# Microscale chemical imaging to characterize and quantify corrosion processes at the metal-electrolyte interface

Cristhiana C. Albert[1], Shishir Mundra[1], Dario Ferreira Sanchez[2], Fabio E. Furcas[1], Ashish D. Rajyaguru[2], O. Burkan Isgor[3], Daniel Grolimund[2], Ueli M. Angst[1,*]

[1]Institute for Building Materials, ETH Zurich, Zurich, Switzerland
[2]Paul Scherrer Institute, Villigen, Switzerland
[3]School of Civil and Construction Engineering, Oregon State University, USA
*corresponding author

**Abstract**

We introduce an experimental setup to chemically image corrosion processes at metal-electrolyte interfaces under stagnant, confined conditions – relevant in a wide range of situations. The setup is based on a glass capillary, in which precipitation of corrosion products in the interfacial aqueous phase can be monitored over time with optical microscopy, and chemically and structurally characterized with microscopic synchrotron-based techniques (X-ray fluorescence, X-ray diffraction, and X-ray absorption spectroscopy). Moreover, quantification of precipitates through X-ray transmission measurements provides in-situ corrosion rates. We illustrate this setup for iron corrosion in a pH 8 electrolyte, revealing the critical role of $O_2$ and iron diffusion in governing the precipitation of ferrihydrite and its transformation to goethite. Corrosion and coupled reactive transport processes can thus be monitored and fundamentally investigated at the metal-electrolyte interface, with micrometer-scale resolution. This capillary setup has potential applications for in-situ corrosion studies of various metals and environments.

## 1. Introduction

Electrochemistry of metals in stagnant electrolytes govern various processes including crevice corrosion [1], corrosion of steel in concrete or soil [2–4], degradation of radioactive waste storage units [5–8], ageing of buried archaeological artefacts [9,10], leaching of medical implants [11,12], or durability and efficiency of batteries [13]. In these conditions, the metal is exposed to a confined electrolyte, where confinement arises from a specific geometry of the solid phases adjacent to the metal, such as crevices or the presence of a porous medium. This confinement entails limited electrolyte renewal at the metal surface, in comparison to a continuously replenishing electrolyte, for instance, streaming seawater. In stagnant electrolytes, transport of ionic or dissolved gaseous species is governed by diffusion and migration, while advection – an important transport process in continuously replenishing electrolytes – is negligible [14].

Despite the importance of stagnant media in a wide range of domains, there is a lack of fundamental understanding, especially regarding the electrochemical and coupled reactive transport processes occurring at the metal-electrolyte interface [2,3,15]. These processes include electrochemical reactions at the metal surface, such as the dissolution of metallic ions in the case of metal corrosion. Species released or consumed in electrochemical reactions undergo diffusion and migration within the electrolyte, as well as chemical reactions [16,17]. The latter includes the oxidation, complexation, and hydrolysis of metal ions, the precipitation of corrosion products, including intermediate phases as well as their transformation to thermodynamically more stable states. It is important to recognize that all these processes are coupled. Thus, reactive transport processes occurring in the stagnant electrolyte near the metal surface affect the kinetics of the electrochemical reactions. Moreover, these reactions as well as the type and distribution of solid corrosion products in the stagnant electrolyte define the stresses potentially exerted to the surrounding media (the solids confining the electrolyte), and their deformation, cracking, and eventual failure [18].

Previous research has shown that experimental investigation of processes in the near field of the metal-electrolyte interface is challenging. Traditional in-situ electrochemical techniques, including benchtop setups such as stirred or unstirred electrochemical cells [19] or with rotating disk electrodes [20,21], can ensure defined hydrodynamic conditions at the metal surface, with diffusion layers of the order of a fraction of a millimeter. However, these setups do not allow to study the location of formation and type of solid reaction products in the electrolyte, as the movement of electrolyte constantly disturbs these processes [19]. Ex-situ surface characterization of corrosion products such as Raman spectroscopy [22,23] or X-ray photoelectron spectroscopy (XPS) [24] focuses on the changes on the metal surface, while less attention is paid to precipitates and their transformations over time in the region of the electrolyte near the metal-electrolyte interface. In-situ synchrotron techniques have also been previously applied in corrosion studies, mainly focusing on pitting corrosion, characterization of surface layers, or atmospheric corrosion [25–28], but rarely to study corrosion products and their evolution along an interface [29–31], or more specifically, in the electrolyte within a confined pore space, as proposed here.

Here, we develop an experimental setup to simulate and study the coupled corrosion and reactive transport processes occurring at the metal-electrolyte interface immersed in confined, stagnant media. The designed setup allows for in-situ characterization, by means of optical microscopy and synchrotron-based chemical imaging of the corrosion products formed, their location, and their undisturbed evolution over time. The high resolution and non-destructive nature of the synchrotron measurements render them ideal techniques for understanding the evolution of corrosion processes, precipitation, and phase transformation of corrosion products at the microscale of the metal-electrolyte interface.

## 2. Capillary setup to characterize the processes at the metal-electrolyte interface

We introduce an experimental setup to investigate in-situ corrosion processes taking place at a metal-electrolyte interface under stagnant and confined conditions (Fig. 1). A metal sample is placed inside a glass capillary that is filled with an electrolyte and is allowed to corrode naturally, i.e., under open-circuit conditions without the application of an external electrical potential or current to accelerate the process. Over time, the processes occurring at the metal-electrolyte interface will lead to the precipitation of solid phases within the capillary and at certain distances from the metal surface [2,8,32]. The precipitates include metastable intermediates as well as thermodynamically more stable phases formed upon transformation of the intermediates. These processes can occur over very different time scales and span over several months to years [33–36]. It is a significant advantage that the proposed setup enables monitoring these precipitation and transformation phenomena in-situ, and thus, in an undisturbed state. Preserving the state during characterization is of particular importance for corrosion products that generally may exhibit different oxidation states, which would potentially be affected upon exposure to air during ex-situ characterization [37–39]. Here, in-situ characterization is achieved through optical and synchrotron-based imaging methods with resolution down to the microscale, providing time-dependent information about the chemical composition, oxidation state, crystalline structure, quantity, and location of the solids formed, as described in more detail below.

Fig. 1 illustrates a situation where two metal-electrolyte interfaces are created. The interface on the left faces the closed end of the capillary, which lacks an $O_2$ source and thus may create an anaerobic environment, while the interface on the right is exposed to the capillary opening, where $O_2$ can enter and establish an aerobic condition. The geometry of the capillary (here, 300 µm diameter and 8 cm in length) together with the location at which the metal is placed within the capillary control the mass-transport conditions, that is, only allow for diffusion and migration, while excluding advection.

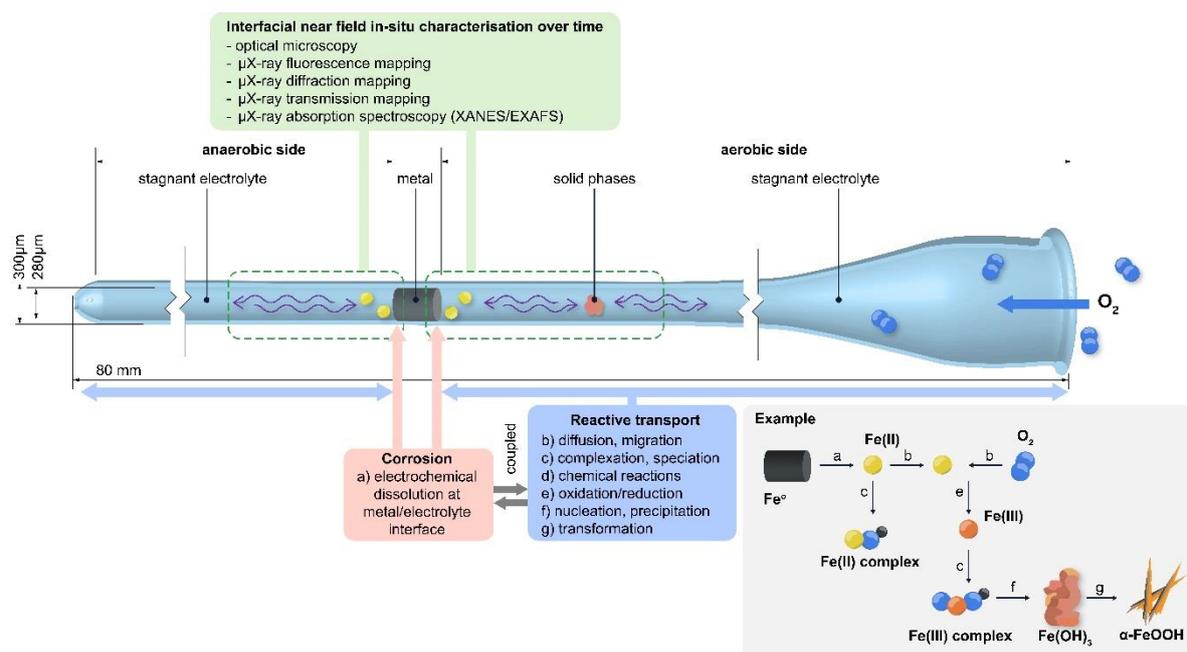

*Fig. 1 – Capillary setup containing a metal and an electrolyte, ensuring controlled mass-transfer processes by maintaining stagnant and confined conditions. The oxygen supply is regulated by adjusting the distance of the metal-electrolyte interface from the capillary's opening on one side. The corrosion process can be understood as a complex and interdependent reactive transport phenomenon, from the electrochemical dissolution of the metal to the precipitation of solid phases (corrosion products) and their transformation, as exemplified for the case of iron. The setup allows for in-situ monitoring of these processes over time, namely with optical and synchrotron-based chemical imaging methods.*

The capillary setup avails a variety of in-situ characterization opportunities. The transparent glass of the capillary enables to monitor the formation and distribution of corrosion products over time by visual inspection and optical microscopy. Moreover, a comprehensive characterization of the corrosion products can be conducted with synchrotron-based techniques. Corrosion processes under open-circuit conditions might lead to the slow precipitation and heterogeneous distribution of corrosion products along the capillary. Therefore, a high-resolution measurement is essential to capture these features at the metal-electrolyte interface and surroundings. Employing a reduced beam size (e.g., full width at half maximum < 2 µm) might allow to measure even early age small precipitates, as conducted here. Since such measurements are non-destructive, the corrosion products formed in the capillary can be characterized over time, providing time-series data that is essential for a thorough investigation of corrosion mechanisms and coupled reactive transport phenomena.

Synchrotron-based chemical imaging methods provide comprehensive and complementary information about the phases formed and their transformation. The chemical composition of the solids can be inspected through μX-ray fluorescence (μXRF) mapping, while the crystalline structure of precipitates can be assessed with μX-ray diffraction (μXRD) mapping. Simultaneously, the μX-ray transmission through the capillary can be recorded and used to quantify the mass of corrosion products. Finally, μX-ray absorption spectroscopy (μXAS) can be used to probe amorphous phases not identified with μXRD. In our case, we performed X-ray absorption near edge structure (XANES) measurements to assess the iron pre-edge feature which contains information on the oxidation state and the coordination environment of the iron in the corrosion products [40–42]. When necessary, a larger energy range can be measured along the extended X-ray absorption fine structure (EXAFS) for more insights into neighbouring atoms [43,44]. However, as XAS experiments are comparatively time consuming, only selected spots are generally measured due to beamtime limitations.

The capillary setup can be adapted for a variety of electrolyte environments and metals. Here, we present an example of high societal relevance: an iron-electrolyte interface, containing an iron wire exposed to an electrolyte of near-neutral pH 8 (5mM Hepes buffer and 1mM $NaHCO_3$). This condition can be considered representative of corrosion of reinforcing steel in carbonated concrete, typically encountered in older infrastructures [2,45], or corrosion of steel in low-alkalinity concrete, which is common with modern, environmentally friendly cements [46]. Moreover, the considered example of iron corrosion in near-neutral pH may also be representative of corrosion of steel in soil, such as in pipelines, tanks, geotechnical applications or archaeological artefacts, or to some extent even steel in bentonite, representative of certain radioactive waste storage approaches [3,9,47–49].

## 3. Results and Discussion

3.1. Evolution of corrosion processes over time

Monitoring the progression of iron corrosion within the capillary with optical microscopy revealed very specific patterns of solid phase precipitation, reproducible across different independently prepared replicate capillaries (Fig. S1). A representative example of this evolution in the aerobic zone of the capillary is shown in Fig. 2, illustrating the precipitation of corrosion products in two regions, one close to the metal surface (II, III) and one further away (V), separated by a zone of almost transparent electrolyte (IV).

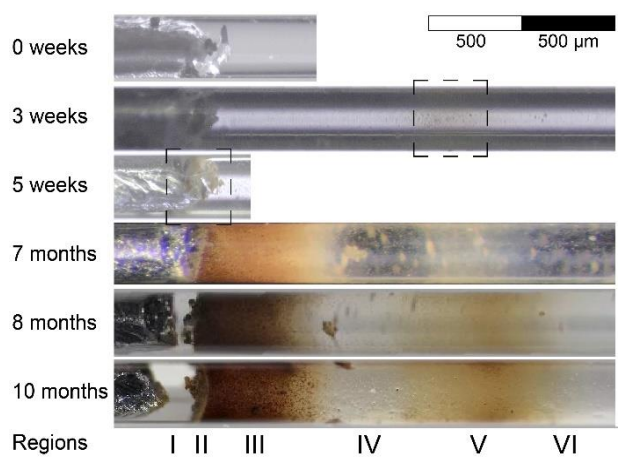

*Fig. 2 – Monitoring of the aerobic side of a capillary using optical microscopy over 10 months, revealing a distinct pattern and evolution of corrosion products precipitation. The dashed boxes indicate the location of precipitation in the early stages.*

In the iron-electrolyte example, the precipitation of corrosion products could only be visually observed in the aerobic side of the capillary (Fig. 1), while the electrolyte in the anaerobic side remained transparent over the entire duration of the experiment (10 months). In the aerobic side, precipitation was first identified after 3 weeks, showing light-yellow coloured corrosion products indicated by the dashed box at region V, more than 1 mm away from the metal (Fig. 2). While this observation was confirmed by replicate capillaries, the exact position of region V depended on the distance between the iron and the capillary opening, indicating that it is affected by oxygen availability. After 5 weeks, light-yellow coloured precipitates were found on the iron-electrolyte interface (dashed box at region II). At 7 months, the regions containing precipitates had grown significantly and evolved into a gradient of brown-orange corrosion products on regions II and III, close to the iron. Further away from the iron, region IV was characterized by a gap containing the transparent electrolyte – indicative of negligible amounts of solid phases – and randomly distributed small precipitates. Region V presented a second gradient of the light-yellow coloured phases, continuously precipitating. At 9 months, the corrosion products were characterized with the synchrotron-based techniques (section 3.2).

In summary, monitoring the capillary with optical microscopy provided the following reproducible observations: (a) there were differences in precipitation patterns between the anaerobic and aerobic areas, (b) the time and location of initial precipitation of corrosion products depended on the distance from the metal surface to the capillary opening (allowing for oxygen supply), (c) distinct regions were developed with varying levels of precipitation, and (d) phases transformed over time, which were associated with changes in colours of precipitates.

3.2. Synchrotron-based characterization

The multimodal synchrotron-based characterization of the corrosion products formed at 9 months ageing yields information on chemistry and structure of the phases, as well as oxidation state and coordination environment of the metal atoms (Fig. 3). These results allow for the identification and quantification of the precipitates along the capillary.

As expected, the distribution of iron atoms shown in the μXRF mapping (Fig. 3b) corresponds to the visual identification of corrosion products via optical microscopy (Fig. 3a). These corrosion products were identified as the Fe(III) hydroxides ferrihydrite ($Fe(OH)_3$) and goethite (α-FeOOH)[1]. Additional details on their characterization are presented in Section 2 of the Supplementary Information. From left to right in Fig. 3, a sequence of regions was observed. First, the iron wire showed its intense signal of iron atoms and crystalline structure (Fig. 3b-c). The iron was followed by the gap on region I with negligible iron atoms and scattering patterns typical of a liquid phase, corresponding well to the transparent electrolyte observed in optical microscopy (Fig. S2). Although the gap regions presented negligible iron signal, they might contain aqueous iron species, but in concentrations too low to be detected by μXRF. Indeed, μXANES conducted on point 2 showed an iron pre-edge feature, mostly fitted as ferrihydrite, while point 1 did not exhibit a well-formed feature resembling an iron pre-edge (Fig. 3e, Fig. S4). This identification of iron hydroxides might have been due to early age small colloidal particles, not yet observable by optical microscopy. Alternatively, such a signal might indicate aqueous species with a similar structure as ferrihydrite, for instance $Fe(OH)_3(aq.)$, which is the most stable Fe(III) species at pH 8 [52].

---

[1] Ferrihydrite ($Fe(OH)_3$) is a metastable and poorly crystalline ferric hydroxide with high surface area [50,51]. 2L-ferrihydrite – 2L-$Fe(OH)_3$, with a bulk composition of $Fe_{9.56}O_{14}(OH)_2$ [51] – is its most amorphous polymorph, and only presents two broad diffraction peaks, making it less distinguishable in XRD analysis. In comparison, 6L-ferrihydrite (6L-$Fe(OH)_3$, also denoted $Fe_{9.86}O_{14}(OH)_2$ [51]) is notably more crystalline, with larger coherent scattering domains, exhibiting six main diffraction peaks. On the other hand, when compared to ferrihydrite, goethite (α-FeOOH) is a more stable phase, associated with lower Gibbs free energy and a crystalline structure.

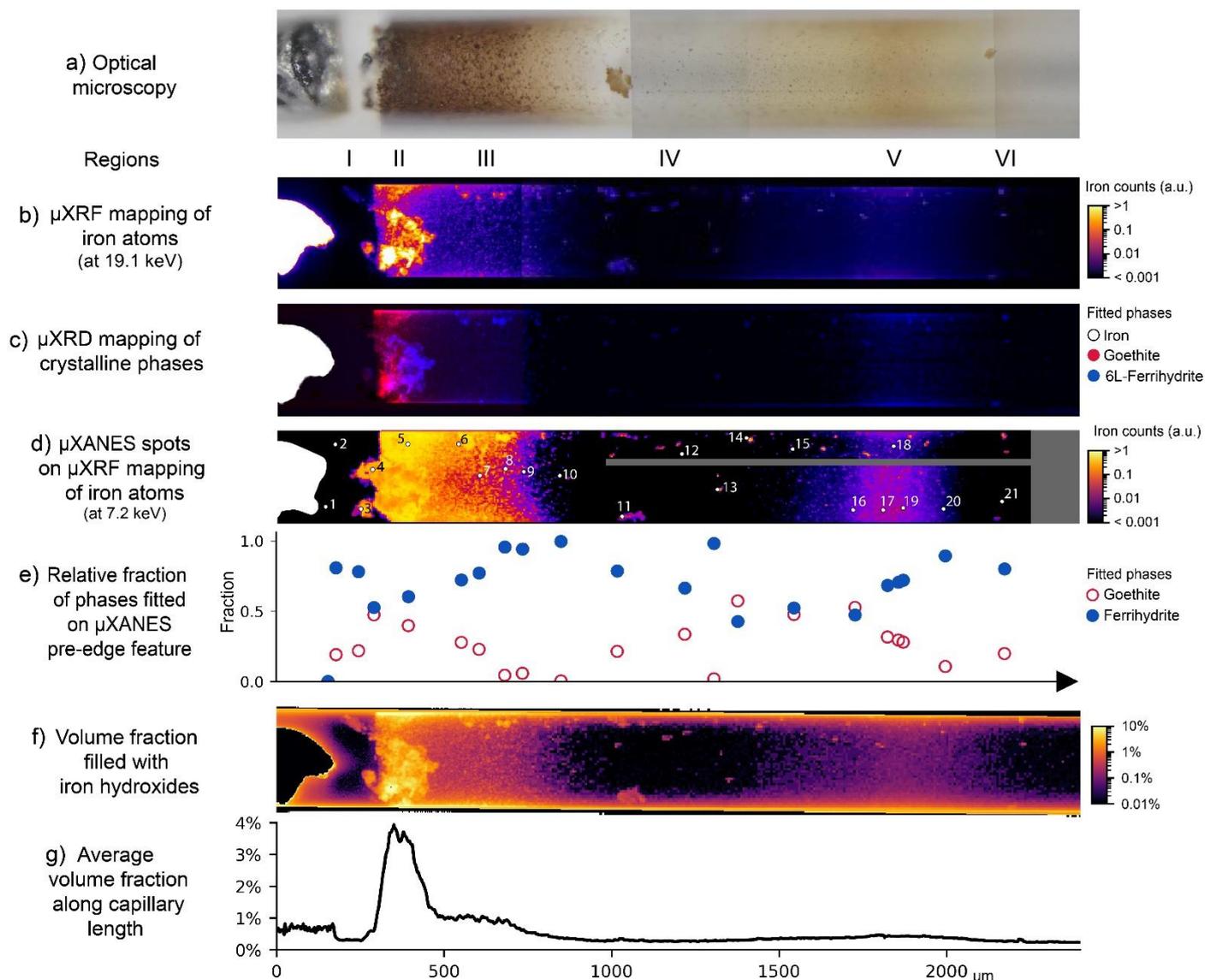

Fig. 3 – *In-situ characterization of corrosion products formed at the iron-electrolyte interface and near-field on the aerobic side of the capillary: (a) optical microscopy at the time of the synchrotron experiments, (b) µXRF mapping of iron atoms, (c) µXRD mapping identifying the distribution of iron (white), goethite (red) and 6L-ferrihydrite (blue), (d) µXRF mapping of iron atoms locating the 21 indicated spots of µXANES measurements, and (e) distribution of goethite and ferrihydrite fractions along the capillary, obtained from linear combination fitting of the XANES pre-edge feature (see suppl. information for more details). Results of X-ray transmission analysis show (f) the capillary volume fraction filled with iron hydroxides, and (g) the average volume fraction filled with iron hydroxides along the length of the capillary.*

The iron counts sharply increased in region II, where the brown precipitates were observed, while along region III the counts gradually decreased, until they became negligible further away from the iron surface (Fig. 3b). The corrosion products in region II were quantified with concentrations up to 1 g/cm$^3$, corresponding to 4% of the capillary volume filled with iron hydroxides – the most concentrated region of the capillary (Fig. 3g). In region III, these average concentrations decreased to 1% of the capillary volume (Fig. 3g). Although the optical microscopy images suggest that corrosion products are dispersed homogeneously within the capillary, the calculated volume fractions are still very low. µXRF and µXRD maps (Fig. 3b-c) show increased iron phases on the top and bottom parts of the images, indicating preferential deposition of corrosion products on the capillary walls. In these zones, the capillary walls are more prominent due to the projection effect arising from representing a 3D cylindrical capillary as a 2D image. This projection artefact may locally overestimate the corrosion products up to 15% at the top and bottom walls of the capillary in Fig. 3g.

Diffraction patterns indicated that region II was mainly composed of goethite, followed by a cluster of 6L-ferrihydrite (Fig. 3c, Fig. S2). Along the gradient in region III, decreasing amounts of goethite and 6L-ferrihydrite were identified with µXRD, with 6L-ferrihydrite primarily precipitating on the walls of the capillary. µXANES confirmed such observations, and points 4 and 5 in region II showed an increasing relative amount of goethite, representing up to ~50% of the corrosion products present there (Fig. 3e, Fig. S4). When progressing along the gradient in region III, the fraction of goethite decreased, reaching its minimum at the end of region III, while ferrihydrite predominated there instead.

The gap region IV generally presented negligible iron counts and scattering patterns (Fig. 3b-c, Fig. S2), in accordance with the very low volume fraction of solids quantified as < 0.01% of the capillary volume (Fig. 3g). µXANES measurements in spots with very low iron counts showed noisy signals and the absence of an iron pre-edge feature, as expected (Fig. S4). Nonetheless, dispersed hot spots of iron were also optically observed as small precipitates, identified mostly as 6L-ferrihydrite according to µXRD (Fig. S4). The complementary results from µXANES, such as point 14 (Fig. S4), indicated that these precipitates are composed of a mixture of ferrihydrite and goethite (Fig. 3e).

Even further away from the iron surface, region V showed a second iron rich region, associated with the light-yellow coloured corrosion products (Fig. 3a). This corresponds to 0.5% of the capillary volume filled with iron hydroxides, an order of magnitude lower than the amount present in region II (Fig. 3g). The diffraction patterns of corrosion products close to the walls of the capillary were mostly amorphous with additional broader peaks (Fig. 3c, Fig. S2). These regions were identified as 6L-ferrihydrite, with a more amorphous nature compared to its appearance closer to the iron, in regions II and III (Fig. S2). No clear signal of goethite were found in region V based on µXRD. Conversely, µXANES showed an increasing fraction of goethite in region V, with its maximum of ~50% at point 16, once again matching the area with higher iron counts (Fig. 3e). This could signify that goethite is present there, possibly in a nano-crystalline state that is not yet detectable with µXRD, but only through µXANES, which measures the local order of the phases [43,44]. Moving to the right side of region V, the relative amount of goethite decreased while the fraction of ferrihydrite increased, such as in the gradient of region III (Fig. 3e). Region VI, apparently without precipitates and with negligible iron counts and crystalline phases, also presented a µXANES spectra with a pre-edge feature similar to ferrihydrite, in point 21 (Fig. S4), as discussed for region I.

In summary, synchrotron-based techniques allow us to comprehensively characterise the evolution of different precipitates formed in the near field of the metal-electrolyte interface. Corrosion products were mostly formed in regions II, III, and V, consisting of goethite and ferrihydrite. The more concentrated and central parts of these regions contained the highest fractions of goethite, more thermodynamically stable than ferrihydrite [33–36,53]. Compared to region II, the further away region V presented a more nanocrystalline goethite only identified with µXANES. These results exemplify the benefit of performing complementary chemical imaging techniques as proposed in our experimental approach. The faster µXRD measurement allows to map a relatively large region in the near field of the metal-electrolyte interface and provides the overall picture, while a closer look at selected spots can be done with the more time-consuming µXANES measurements. Moreover, while the X-ray transmission approach might locally overestimate corrosion products due to an artefact of the projected capillary walls, µXRF and µXRD directly probe the iron phases and are not influenced by these edge effects.

3.3. In-situ corrosion rates

Based on the amount of iron corrosion products formed in the aerobic side of the capillary at 9 months (Fig. 3), the averaged corrosion rate at that iron-electrolyte interface was calculated as 0.8 µA/cm$^2$, approximately equivalent to 10 µm/year. This value is of the same order of magnitude of long-term average corrosion rates of environments similar to the iron-electrolyte capillary, such as steel in wet carbonated concrete, which is typically 5-20 µm/year [54], or long-term average corrosion rates of buried iron archaeological artefacts, under 4 µm/year [9].

The anaerobic side of the capillary did not exhibit visible corrosion products after 10 months. Thus, we conclude that the maximum mass of dissolved iron cannot have significantly surpassed the solubility limit of Fe(II) and Fe(III) species. Based on this solubility limit, a maximum corrosion rate at the anaerobic iron-electrolyte interface was estimated as 0.03 µA/cm$^2$ (0.4 µm/year). This is at least 30 times lower than the corrosion rate calculated for the aerobic iron-electrolyte interface. Such results are in agreement with reported anaerobic corrosion rates for steel in bentonite at near-neutral pH, such as in radioactive waste repositories, reported to be around 0.2 µm/year [7,55].

In the synchrotron-based chemical imaging experiments presented here, the corrosion of iron inside the capillary was characterized in-situ with negligible external disturbance. The X-ray energies and fast measurements are not considered to modify the natural processes taking place inside the capillary, as confirmed by the similar XANES spectra obtained in subsequential measurements in the same spot. Thus, the approach proposed here allows for the determination of undisturbed corrosion rates. We consider this an advantage over conventional experimental setups, where corrosion rates are commonly determined using gravimetric or electrochemical measurements. Although these are valuable techniques in corrosion science, gravimetric methods involve errors that originate from subjective corrosion product removal and weighing errors, while electrochemical methods are susceptible to errors associated with experimental setup (e.g., wiring, sample holder, masking, contact resistance), test parameter selection (e.g., applied potential or current, scan range and rate), and data interpretation (e.g., fitting of the data) [56].

We emphasize the potential of the proposed X-ray transmission approach to quantify the in-situ corrosion processes in confined stagnating conditions in a time-resolved manner. Such corrosion processes occur in pores, where it can be challenging to determine corrosion rates with classical techniques. The proposed approach is also particularly valuable for slow electrochemical processes, which are hardly quantifiable with electrochemical and gravimetric techniques [56].

3.4. Perspective on corrosion mechanism at the metal-electrolyte interface

The experimental approach presented here monitors the in-situ evolution of corrosion products, by means of optical microscopy, with the synchrotron-based characterization of the phases, in terms of chemical composition, structure, oxidation state and coordination environment of the metal in the corrosion products. Such comprehensive information is highly valuable to gain mechanistic insights on metal corrosion and the processes occurring at the metal-electrolyte interface. To illustrate this and give a perspective on how the acquired data may be harnessed to elucidate fundamental mechanisms, we briefly discuss the example of iron corrosion in near-neutral media below.

Corrosion products were observed to form only in the aerobic side of the capillary and distributed in two main zones. Initially, solid phases precipitated further away (region V) and, later, closer to the iron (regions II, III). Their colours changed over time, suggesting a phase transformation process. At 9 months, these precipitates were characterized as two types of Fe(III) hydroxides: goethite – mainly found in the regions of the capillary with higher amounts of corrosion products – and ferrihydrite. These observations can be rationalized with thermodynamic and kinetic data, showing that ferrihydrite is a metastable phase that transforms to the more crystalline and stable phases goethite and hematite, where the transformation rate is a function of the pH, temperature and the presence and concentration of additional solutes [33–36,53]. Based on these observations, we can infer a sequence of precipitation and ageing processes in the capillary.

Accordingly, corrosion products precipitated first as ferrihydrite, with the highest solubility among all characterized Fe(III) phases [52], and then, over time, transformed to goethite [33–36,53]. An interesting observation in our experiment was, that even though the corrosion products in region II precipitated in a later stage, their phase transformation to goethite was more advanced compared to the corrosion products in region V.

In addition, we highlight the critical role of $O_2$ for the precipitation of corrosion products in very specific patterns along the aerobic side of capillary. When iron corrosion initiates, aqueous Fe (II) species form and diffuse along the length of the capillary. Fe(II) hydroxides can precipitate if the local concentration of Fe(II) surpasses its solubility limit. Such Fe(II) precipitation is, however, unlikely in the here studied case due to the relatively high solubility of $10^{-3}$ molL$^{-1}$ for Fe(II) w.r.t. Fe(OH)$_2$ (s) at pH 8 [57]. Alternatively, Fe(II) can oxidize to Fe(III), which has a significantly lower solubility limit of only $10^{-9}$ molL$^{-1}$ w.r.t. Fe(OH)$_3$ (s) at pH 8 [57]. Thus, once oxidized, the precipitation of Fe(III) hydroxides is facilitated. The oxidation rate of Fe(II) to Fe(III) depends on pH and the concentrations of Fe(II) and $O_2$ in solution – e.g., the oxidation reaction is 10 times faster in aerobic than in anaerobic conditions at pH 8 [57].

Moreover, the presence of $O_2$ at the metal surface accelerates the anodic dissolution of iron and increases the Fe(II) concentration in solution, thanks to the reduction of $O_2$ as the primary cathodic reaction. This accelerated release of Fe(II) promotes the precipitation of corrosion products according to the mechanism discussed on the previous paragraph. On the other hand, the electrochemical dissolution of Fe(II) on the anaerobic side of the capillary is sustained by the cathodic reaction of $H_2$ evolution. Due to the limited overpotential when interacting with anodic iron dissolution at pH 8, the $H_2$ evolution reaction is slow, which explains the at least 30 times lower corrosion rate calculated for the anaerobic metal-electrolyte interface compared to the aerobic one. In summary, the precipitation of Fe(III) hydroxides is highly favoured in aerobic electrolytes at near-neutral pH, while in anaerobic conditions, most ions probably remain as aqueous Fe(II) or Fe(III) species at concentrations below their respective solubility limits. If formed, precipitates are present in low amounts and not yet observable by optical microscopy.

The capillary setup can thus provide information on the location, type, and transformation of precipitated phases. Combining these observations with knowledge on thermodynamic stability and reaction kinetics (e.g. Fe(II) → Fe(III) oxidation or phase transformation rates) allows for inferring the local chemistry of the electrolyte. Time-resolved experiments might further shed light on aspects such as the dynamic corrosion rate of a metal in a pore, diffusion coefficients of aqueous metallic species and $O_2$, the influence of $O_2$ on the oxidation of ionic metal species, and the precipitation and transformation of solid phases (corrosion products). These corrosion processes may be seen as a reactive transport phenomenon, as schematically illustrated in Fig. 1. We consider it a particular advantage that the comprehensive experimental data obtained from this capillary setup can thus support and validate numerical models describing this reactive transport phenomenon [16,17]. In particular, experimental quantification of the time and location of precipitation as well as the identification of corrosion products and their transformation can support the validation of intermediate steps of reactive transport models, which currently – in the absence of such data – is difficult.

In conclusion, the proposed capillary setup offers a unique opportunity to investigate coupled corrosion and reactive transport processes at the metal-electrolyte interface within a single specimen under realistic conditions, which is not possible with standard bench top electrochemical experiments. The range of metal-electrolyte interfaces to be explored includes a variety of solution environments, different metals, as well as the addition of particles to the electrolyte to simulate porous media such as concrete, soil, or bentonite. Employing special capillaries (e.g. sapphire capillaries) can also allow the investigation of high-temperature and/or high-pressure conditions, relevant for nuclear waste disposal for example [7,8,55]. Further characterization could be conducted with synchrotron based chemical microtomography for the 3D evaluation of the distribution of precipitates, such as in opaque porous media, where the corrosion products cannot be easily distinguished by optical microscopy [38].

## 4. Methods

4.1. Example of steel in near-neutral electrolyte

The setup consists of a borosilicate glass capillary (300 μm external diameter, 10 μm wall thickness, and 8 cm length) containing an iron-electrolyte interface, as schematically illustrated in Fig. 1. We exemplify an application of such a setup by mimicking an iron-electrolyte interface. The capillary was first saturated with a solution buffered at pH 8 (5 mmolL$^{-1}$ HEPES buffer and 1 mmolL$^{-1}$ NaHCO$_3$), simulating environments such as carbonated concrete or a variety of soil types. A 99.99% iron wire (250 μm diameter, < 1 mm length) was then inserted in the capillary, creating two interfaces between the iron and solution, as a simplified model for a metal-electrolyte interface. The capillary was stored in the vertical position in a closed container filled with the pH 8 solution, at room temperature ~23°C and atmospheric pressure. The bottom of the capillary is closed (left side in Fig. 1) while the top of the capillary is open (right side in Fig. 1), exposing each side of the iron surface to an anaerobic and aerobic environment, respectively. The evolution of the corrosion products inside the capillary was regularly monitored by means of optical microscopy and, at 262 days, the capillary was subjected to the synchrotron experiments, as described in the next section.

4.2. In-situ characterization over time

  4.2.1. Optical microscopy

The precipitation of corrosion products inside the capillary was monitored by means of optical microscopy over 10 months, applying magnifications of 40x and 100x. It should be noted that the cylindrical shape of the capillary, with an angular surface along the y axis, might lead to irregular light refraction which could induce distortions on the recorded image. These distortions were corrected based on the chemical imaging by μXRF mapping.

  4.2.2. μX-ray fluorescence mapping

The synchrotron experiments were conducted at the microXAS beamline, a microprobe facility at the SwissLight Source (SLS) hosted by Paul Scherrer Institut (PSI, Switzerland). The distribution of elements in the iron-electrolyte interface and surrounding region were 2D mapped with µX-ray fluorescence (µXRF). A solid-state fluorescence detector (Ketek GmbH, Germany) with an incident X-ray energy of 19.1 keV, above the FeK absorption edge and high enough for µXRD acquisition. The capillary was mapped in two parts using the on-the-fly scanning mode: scan 1, closer to the iron (x < 750 µm), was measured with step size of 2.5 µm, whereas scan 2 (750 µm < x < 2400 µm) was conducted with step size of 7.5 µm. These step sizes were applied for the diffraction and transmission mappings as well. The µXRF data were processed using PyMCA [58].

### 4.2.3. µX-ray diffraction mapping

Simultaneously with the µXRF mapping, the capillaries were 2D mapped with µX-ray diffraction (µXRD). Diffraction patterns were obtained using a fast 2D detector (Eiger 4M, DECTRIS ltd, Switzerland), with an incident X-ray energy of 19.1 keV ($\lambda$ = 0.6491 Å). The data was analysed with the XRDUA software [59], with a background removal in the scan 2 by subtracting the average amorphous halo obtained from the regions of the capillary only containing the solution and negligible iron counts. The peaks were fitted with the powder diffraction standards of iron, goethite, and 6L-ferrihydrite (ICSD entries 24963, 245057, 158475, respectively). More details are found in Fig. S2.

The colour intensity at each pixel is calculated from the height of the main XRD peaks fitted for each phase. These results are semi-quantitative because the exact phase composition at each pixel cannot be quantified, both due to the cylindrical shape of the capillary and due to the possible presence of amorphous phases that are not accounted for in XRD.

### 4.2.4. µXANES

µX-ray absorption spectroscopy (XAS) measurements covering the XANES region were conducted on selected points along the capillary, in order to confirm the findings of µXRD and characterize amorphous phases. First, µXRF mapping was done at 7.2 keV, indicating iron rich areas and gradients from which spots were selected to represent diverse regions with low and high iron counts. µXANES was recorded in the fluorescence mode between 7.05 – 7.5 keV, including the Fe-K pre-edge feature, measured with resolution of 0.25 eV, and with Fe-K edge calibration at 7.112 keV, according to [60]. A Si (111) monochromator was used, and the beam size was 1.5 µm. The data were processed using the Athena software, including normalization and self-absorption correction based on transmission mode measurements. The similarity between the pre-edge features of 2L-ferrihydrite and 6L-ferrihydrite limits to reliably distinguish them from each other [61]. Therefore, in this analysis we describe both as a single ferrihydrite. The fractions of goethite and ferrihydrite were determined by means of pre-edge linear combination fitting, applying standards measured at the superXAS beamline of SLS [62], and considering an energy shift of -0.17 eV. The pre-edge background was fitted using a rubberband baseline, which minimized the influence of the chosen fitting range in comparison to the spline method. Gaussian peaks were used to fit the relevant pre-edge region of the standard spectra, constraining the distance between peaks based on the literature and also adding a third peak to ferrihydrite [40,63], as reported on Table 1 and shown in Fig. S3. Examples of peak fitting are shown in the Fig. S4.

Table 1 – Fitting parameters of goethite and ferrihydrite standards

|  | Center (eV) | FWHM | Amplitude |
|---|---|---|---|
| Goethite peak 1 | 7113.6 | 1.437 | 0.01566 |
| Goethite peak 2 | 7115.0 | 1.682 | 0.03192 |
| Ferrihydrite peak 1 | 7113.9 | 1.790 | 0.04922 |
| Ferrihydrite peak 2 | 7114.9 | 1.765 | 0.03948 |
| Ferrihydrite peak 3 | 7116.6 | 1.684 | 0.00559 |

### 4.2.5. μX-ray transmission mapping

The transmitted beam intensity after the beam attenuation through the capillary was also collected during the μXRF and μXRD 2D mapping, shown in Fig S5. Based on the Beer-Lambert Law, the attenuation was calculated as $\mu z = -\ln(I_o/I)$, considering $\mu$ as the average attenuation coefficient of the capillary, $z$ as the depth of the capillary at each point, and the incident beam energy of $I_o$ = 19.1 keV. Tabulated values of the linear attenuation coefficients of glass ($\mu_{SiO2}$) and corrosion products ($\mu_{CP}$) were used for the analysis. The solution does not contribute to beam attenuation at the energy level employed and was not considered.

First, the empty regions of the capillary were analysed with the $\mu_{SiO2}$ at 19.1 keV, resulting in an average depth of 20 μm associated with the glass, in accordance with the thickness of the capillary walls, and thus validating the method. It should be noted that, due to the 3D cylindrical shape of the capillary, its projection into a 2D image creates an artefact of thicker walls on the edges – top and bottom parts of the image. This means that a part of the attenuation measured at these regions is due to the additional contribution of the glass, and thus the iron hydroxide amount is overestimated on the top and bottom parts of Fig. 3f. This overestimation is at most 15% at the spot with maximum wall thickness in the 2D projection.

The $\mu z$ contribution of the glass was subtracted from the entire capillary, such that the remaining $\mu z$ should be associated with the iron and corrosion products. The average $\mu$ of goethite, 2L-ferrihydrite, and 6L-ferrihydrite was assumed ($\mu_{CP}$ = 9.1 ± 0.8 mm$^{-1}$), with a reasonable standard deviation below 9%. Thus, the average depth of the iron hydroxides was calculated at each point and, considering the spatial resolution at each pixel, a volume and weight of iron hydroxides was found. By additionally calculating the full depth of the capillary at each pixel, the iron hydroxides mass/volume and volume fraction of the capillary filled with iron hydroxides were quantified.

### 4.2.6. Corrosion rate estimation

The corrosion rates are estimated for the aerobic and anaerobic iron-electrolyte interfaces (Fig. 1), considering that they undergo independent corrosion processes and that the capillary is confined and does not allow the exit of iron species or solids.

The X-ray transmission mapping provides the mass of corrosion products in the aerobic side of the capillary. This mass was converted into the average mass of iron atoms present in the corrosion products (in our case, goethite and ferrihydrite, as characterized with the μXRD and μXANES experiments), resulting in the mass of iron that corroded over the time of the experiment (261 days). Based on Faraday's Law of Electrolysis, the average corrosion rate of iron was calculated for the iron-electrolyte interface exposed to the aerobic conditions.

Since corrosion products were not visually observed in the anaerobic side of the capillary, we assumed that all iron phases remain as soluble Fe(II) or Fe(III) species. Considering their solubility limits at pH 8 ($10^{-3}$ molL$^{-1}$ and $10^{-9}$ molL$^{-1}$, respectively) and the volume of the anaerobic zone of the capillary, the maximum mass of iron that could have corroded was calculated. Faraday's Law of Electrolysis was then used to calculate the maximum corrosion rate for the iron-electrolyte interface exposed to anaerobic conditions.

**ACKNOWLEDGEMENTS**

The authors are grateful to the European Research Council (ERC) for the financial support provided under the European Union's Horizon 2020 research and innovation programme (grant agreement no. 848794). A.R. was supported by the European Union's Horizon 2020 research program under the Marie Sklodowska-Curie (grant agreement no. 701647) and partly by the PSI internal funding scheme-Cross Project. We acknowledge Dr. Emanuele Rossi, Dr. Mohit Pundir, Xiulin Chen, and the staff of MicroXAS at SLS for experimental assistance.

## AUTHOR CONTRIBUTIONS

The study was conceived by S.M., A.R., D.G., U.A., and C.A.. A.R. developed the capillary setup. C.A. prepared the materials and C.A., S.M., F.F. and D.F.S. performed the experiments. C.A. and U.A. analysed the data under the supervision of S.M., D.F.S., and D.G., with advice of B.I.. All authors discussed the data interpretation. C.A. wrote the paper's first draft, revised by all authors. All authors approved the final version.

## COMPETING INTERESTS

The authors declare no competing interests.

## ADDITIONAL INFORMATION

**Supplementary information** The online version contains supplementary material available at xxx
**Correspondence** and requests for materials should be addressed to U. Angst.